\documentclass[journal=nalefd,manuscript=letter]{achemso}

\usepackage{multicol}

\usepackage{xcolor}
\usepackage{hyperref}

\usepackage[version=3]{mhchem} % Formula subscripts using \ce{}

\author{Tom\'as Santiago-Cruz}
\affiliation[MPL]
{Max Planck Institute for the Science of Light, Staudtstra{\ss}e 2, 91058 Erlangen, Germany.}
\alsoaffiliation[FAU]
{University of Erlangen-N\"urnberg, Staudtstra{\ss}e 7/B2, 91058 Erlangen, Germany.}
\alsoaffiliation[MPSP]
{Max Planck School of Photonics, Albert-Einstein-Str. 6, 07745 Jena, Germany.}
\altaffiliation{These authors contributed equally.}
\email{jose-tomas.santiago@mpl.mpg.de}
\author{Anna Fedotova}
\affiliation[Jena]
{Institute of Applied Physics, Abbe Center of Photonics, Friedrich Schiller University Jena, 07745 Jena, Germany.}
\altaffiliation{These authors contributed equally.}
\author{Vitaliy Sultanov}
\affiliation[MPL]
{Max Planck Institute for the Science of Light, Staudtstra{\ss}e 2, 91058 Erlangen, Germany.}
\alsoaffiliation[FAU]
{University of Erlangen-N\"urnberg, Staudtstra{\ss}e 7/B2, 91058 Erlangen, Germany.}
\author{Maximilian A. Weissflog}
\affiliation[Jena]
{Institute of Applied Physics, Abbe Center of Photonics, Friedrich Schiller University Jena, 07745 Jena, Germany.}
\alsoaffiliation[MPSP]
{Max Planck School of Photonics, Albert-Einstein-Str. 6, 07745 Jena, Germany.}
\author{Dennis Arslan}
\affiliation[Jena]
{Institute of Applied Physics, Abbe Center of Photonics, Friedrich Schiller University Jena, 07745 Jena, Germany.}
\author{Mohammadreza Younesi}
\affiliation[Jena]
{Institute of Applied Physics, Abbe Center of Photonics, Friedrich Schiller University Jena, 07745 Jena, Germany.}
\author{Thomas Pertsch}
\affiliation[Jena]
{Institute of Applied Physics, Abbe Center of Photonics, Friedrich Schiller University Jena, 07745 Jena, Germany.}
\alsoaffiliation[MPSP]
{Max Planck School of Photonics, Albert-Einstein-Str. 6, 07745 Jena, Germany.}
\alsoaffiliation[IOF]{Fraunhofer Institute for Applied Optics and Precision Engineering, 07745 Jena, Germany.}
\author{Isabelle Staude}
\affiliation[Jena]
{Institute of Applied Physics, Abbe Center of Photonics, Friedrich Schiller University Jena, 07745 Jena, Germany.}
\alsoaffiliation[JenaSolidState]
{Institute of Solid State Physics, Friedrich Schiller University Jena, 07743 Jena, Germany.}
\author{Frank Setzpfandt}
\affiliation[Jena]
{Institute of Applied Physics, Abbe Center of Photonics, Friedrich Schiller University Jena, 07745 Jena, Germany.}
\author{Maria V. Chekhova}
\affiliation[MPL]
{Max Planck Institute for the Science of Light, Staudtstra{\ss}e 2, 91058 Erlangen, Germany.}
\alsoaffiliation[FAU]
{University of Erlangen-N\"urnberg, Staudtstra{\ss}e 7/B2, 91058 Erlangen, Germany.}
\alsoaffiliation[MPSP]
{Max Planck School of Photonics, Albert-Einstein-Str. 6, 07745 Jena, Germany.}

\title[An \textsf{achemso} demo]
  {Spontaneous Parametric Down-Conversion from Resonant Metasurfaces}

\abbreviations{IR,NMR,UV,FP}

\begin{document}

\begin{abstract}
All-dielectric optical metasurfaces are a workhorse in nano-optics due to both their ability to manipulate light in different degrees of freedom and their excellent performance at light frequency conversion. Here, we demonstrate first-time  generation of photon pairs via spontaneous parametric-down conversion in lithium niobate quantum optical metasurfaces with electric and magnetic Mie-like resonances at various wavelengths. By engineering the quantum optical metasurface, we tailor the photon-pair spectrum in a controlled way. Within a narrow bandwidth around the resonance, the rate of pair production is enhanced up to two orders of magnitude compared to an unpatterned film of the same thickness and material. These results enable flat-optics sources of entangled photons -- a new promising platform for quantum optics experiments.
\end{abstract}

%\begin{multicols}{2}

%############################### Introduction ###########################

\section{Introduction}

An important current tendency is the miniaturization of photonic devices towards multifunctional films with thicknesses in the nanometer range \cite{yu:2014:NM}, so-called metasurfaces. Metasurfaces already found several applications in linear optics, like ultrathin lenses \cite{khorasaninejad:2017:S}, holograms \cite{zheng:2015:NN}, mode converters  \cite{sroor:2020:NP}, and filter structures \cite{berzins:2019:ACSP}. Recently, they have also been established as a promising platform for nonlinear optics \cite{zou:2019:JoPD}. The advantages of nonlinear flat-optics devices are their slim profile, ultrafast and broadband operation, and relaxed phase matching in frequency conversion~\cite{Liu2018}. The latter leads to unprecedented freedom in the choice and engineering of nonlinear materials. Nanoscale sources with engineered nonlinearities and resonant field enhancement provide efficiencies of nonlinear frequency conversion as high as $1\%$~\cite{Krasnok2018} and versatile nonlinear beam shaping \cite{gao:2018:NL}. However, most of the previous work has been focused on the generation and control of classical light. The next natural step are quantum optical metasurfaces (QOM) for the generation of quantum states of light through nonlinear processes at the nanoscale~\cite{Solntsev2020,poddubny2020quantum}.

While single-photon emitters have already been integrated into flat-optics platforms, the efficient generation of photon pairs within such structures remains a challenge. Pair generation has been reported using four-wave mixing in carbon nanotube films~\cite{Lee2017}, but the efficiency and signal-to-noise ratio were low. Another commonly used nonlinear process to create photon pairs is spontaneous parametric down-conversion~(SPDC). In SPDC, a pump photon of higher frequency $\omega_p$ splits in two daughter photons, signal and idler, with lower frequencies $\omega_s$ and $\omega_i$, where energy conservation requires $\omega_s+\omega_i=\omega_p$. %On nanoscale subwavelength thickness of the source relaxes the phase matching condition and grants the unprecedented freedom in engineering of nonlinear materials. 
Recently, photon-pair generation by SPDC was demonstrated in ultrathin films of lithium niobate (LN) and gallium phosphide (GaP)~\cite{Okoth2019,Santiago2020}. But despite the high bulk second-order susceptibilities $\chi^{(2)}$ of these materials, the achieved pair generation rates were modest: SPDC is based on the parametric amplification of the vacuum field~\cite{klyshkoNonlinearoptics}, which is extremely weak. Furthermore, contrary to classical nonlinear processes like second-harmonic generation (SHG), the SPDC rate scales linearly with the pump power, and hence its efficiency cannot be enhanced by using a pulsed or focused pump. 

To boost the efficiency of nanoscale SPDC, nanostructures with resonances at the signal and idler frequencies can be employed. Such structures feature an increased density of states, thus enhancing vacuum fluctuations and enabling more efficient SPDC. This approach was already studied theoretically \cite{Poddubny2018,nikolaeva:2020:arxiv} and experimentally \cite{marino:2019:O} for single nanoresonators exhibiting Mie-type resonances. Although the demonstrated results are promising, they are limited by the small total volume of the used resonators. Nonlinear metasurfaces -- two-dimensional arrangements of such nonlinear nanoresonators -- promise a better photon-pair generation rate. 

Especially all-dielectric high-$\chi^{(2)}$ metasurfaces~\cite{Pertsch2020} 
are a favourable platform for SPDC, thanks to their high damage threshold. Various resonance effects, such as Mie-type~\cite{Liu2016,Gili2016} or Fano-type~\cite{Vabishchevich2018} resonances and bound states in the continuum~\cite{Koshelev2019}, have been shown to enhance SHG efficiency by several orders of magnitude~\cite{Vabishchevich2018,Anthur2020}. Due to the similarities between classical parametric frequency conversion and SPDC \cite{liscidini:2013:PRL,helt:2015:OL,lenzini:2018:LSA}, comparable enhancements are expected for the latter.
%%%%

Here we observe, for the first time, SPDC from resonant QOMs, schematically shown in Fig.~\ref{fgr:metasurface_design}(a).  
Due to the resonances, the photon pairs are emitted only into a narrow wavelength range, which opens a possibility to engineer their spectrum. 
Furthermore, within the emission bandwidth we observe two orders of magnitude enhancement of the pair generation rate compared to an unstructured film of the same thickness as the metasurface, despite the fact that nanostructuring reduces the volume of the nonlinear material and our optics do not collect all photon pairs generated from the QOM. The experiment is run in the reflection geometry, common for nonlinear optics and fluorescence experiments, but quite `unorthodox' for SPDC.

%%%%%%%%%%%%%%%%%%
%############################### Results and discussion ###########################

\section{Results and discussion}

\begin{figure}[t!]
     \includegraphics[width=16cm]{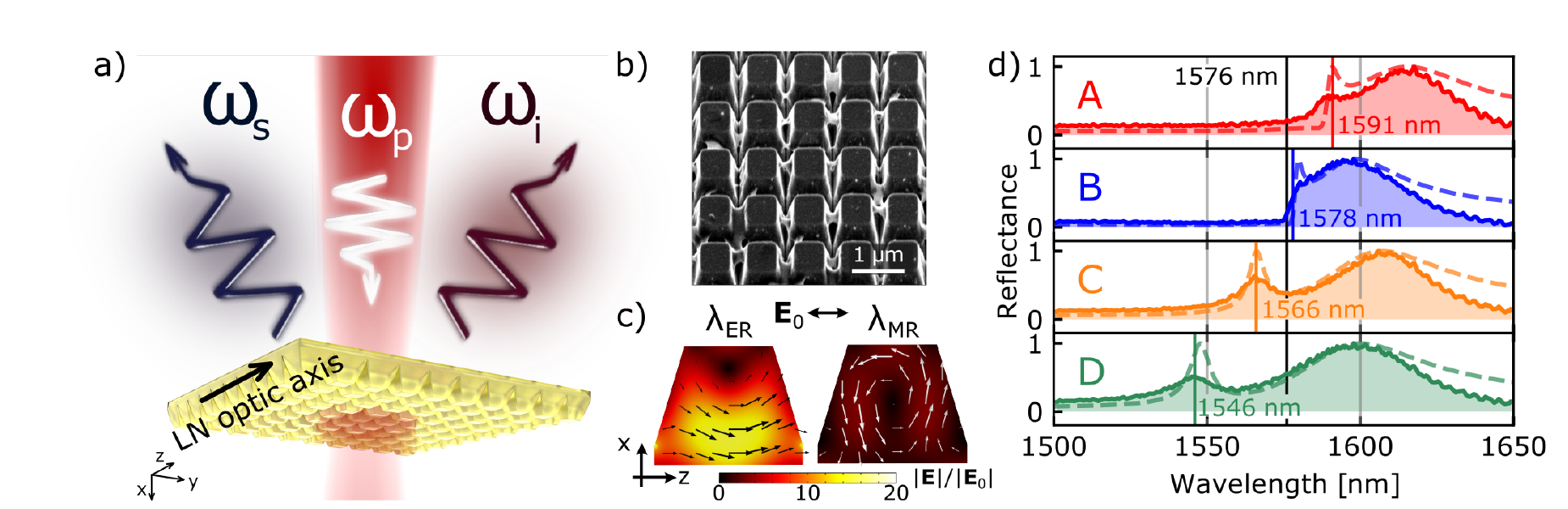}
  \caption{(a) Artist's view of SPDC from a LN metasurface: the pump is incident from the substrate side, photon pairs are collected in reflection. Both the pump and the SPDC photons are polarized along the LN optic axis $z$. (b) Scanning electron microscope~(SEM) image of a fabricated metasurface showing a periodic array of nanoresonators in the shape of truncated pyramids. (c) Electric field distribution inside such a nanoresonator as calculated in COMSOL Multiphysics at the electric~(left) and magnetic~(right) resonances. The incident field is polarized along the LN optic axis. Arrows show the electric field direction. (d) Experimental~(solid lines, shading) and simulated~(dashed lines) reflectance spectra of four QOMs with different resonance positions, further labeled as A, B, C, D; vertical black line marks the wavelength of degenerate SPDC. Vertical colored lines mark the positions of the electric resonances.}
  \label{fgr:metasurface_design}
\end{figure}	

	%We have fabricated metasurfaces on a 680 nm thick LN film on insulator. The metasurfaces consist of nanoresonators in the shape of truncated pyramids with side lengths around 700 nm and periods around 900 nm, so as to sustain Mie-like resonances in the telecom wavelength range. Panel b of Figure xx shows a scanning electron micrograph of one metasurface. 

	%Using a custom-built white-light spectroscopy setup described elsewhere~\cite{Fedotova2020}, we measured the white-light transmittance of two metasurfaces with side length of 700 nm and period of 890 nm  and  930 nm. The results are shown by the blue and red curves in Figure xx.c, respectively. Both metasurfaces exhibit a narrow resonance and a broad resonance as shown by the two dips. Important here, the reflectance spectra would show two peaks at the resonant wavelengths in the absence of absorption. To support our observations, we carried out numerical simulations  using the finite element method using the software COMSOL Multiphysics. The results of our simulations are shown by dashed lines in Figure xx.c. 
%%%%%%%%%%%%%%%%%	

%%%%%FS version%%%%%%
We generate photon pairs in QOMs made of LN, which is well known for its high second-order susceptibility $\chi^{(2)}$. Several recent works already demonstrated efficient SHG with LN nanoparticles~\cite{Timpu2019} and metasurfaces~\cite{gao2019lithium,Ma2020,carletti:2021:ACSP,Fedotova2020}. 
LN is especially attractive for quantum nonlinear optics due to its broad transparency range and relatively low fluorescence compared to semiconductors like GaAs or GaP. In this work we take advantage of its largest nonlinear tensor component $\chi^{(2)}_{zzz}$.

Our metasurfaces are designed with fundamental magnetic and electric resonances for signal and idler photons. At resonance, the density of states is increased, enabling enhanced generation rates for photon pairs \cite{helt:2012:JOSAB}. A resonance at the pump wavelength could further increase the SPDC rate; however, since SPDC depends linearly on the pump power, the same effect can be achieved by modifying the pump properties. 

We have fabricated QOMs on a 680~nm thick x-cut LN-on-insulator film. The metasurfaces consist of nanoresonators in the shape of truncated pyramids with side lengths around 700~nm, arranged with a period around 900~nm as shown in a scanning electron micrograph in Fig.~\ref{fgr:metasurface_design}(b).
Our QOMs support two Mie-like resonances in the near-infrared wavelength range~\cite{Fedotova2020}, further called `electric' and `magnetic' for the reasons discussed below. Their field distributions inside one nanoresonator are shown in Fig.~\ref{fgr:metasurface_design}(c), left and right panels, correspondingly.    

Using a custom-built white-light spectroscopy setup, 
we measured the reflectance of our QOMs for light polarized along the optic axis of LN (Fig.~\ref{fgr:metasurface_design}(d), solid lines). Each QOM exhibits a narrow electric resonance at shorter and a broad magnetic resonance at longer wavelengths, indicated by the maxima in the reflection spectra. Four QOMs have been investigated, with the following wavelengths of electric resonances: 1591~nm (A), 1578~nm (B), 1566~nm (C), and 1546~nm (D). The geometric parameters and the resonance wavelengths of all QOMs are listed in the Supporting Information (SI).
With reducing the size/period of the pyramids, the resonances shift towards shorter wavelengths. Our experimental observations are corroborated by numerical simulations with the finite element method (dashed lines in Fig.~\ref{fgr:metasurface_design}(d)). 
The short-wavelength electric resonance, although featuring several multipole components, is dominated by the electric dipole and quadrupole with the electric fields mainly in the plane of the metasurface~(see the black arrows in Fig.~\ref{fgr:metasurface_design}(c)). The long-wavelength magnetic resonance features electric field with a more complex structure and is dominated by the magnetic dipole. More details on the resonances can be found in the SI.

Classical frequency-conversion experiments showed that electric-type resonances with fields in the metasurface plane can fully utilize the strongest component of the LN second-order susceptibility tensor $\chi^{(2)}_{zzz}$, leading to improved conversion efficiencies~\cite{Fedotova2020}. Furthermore, because the electric resonances in our metasurfaces have higher field enhancement than the magnetic resonances~(see Fig.~\ref{fgr:metasurface_design}(c) and the SI),  they also provide a larger enhancement of the density of states. Therefore, for our SPDC experiments pumped at 788~nm we chose QOMs with electric resonances near the degenerate photon-pair wavelength $\lambda_{\text{deg}}=2\cdot788$~nm$=1576$~nm.

\begin{figure}[t!]
     \includegraphics[width=15cm]{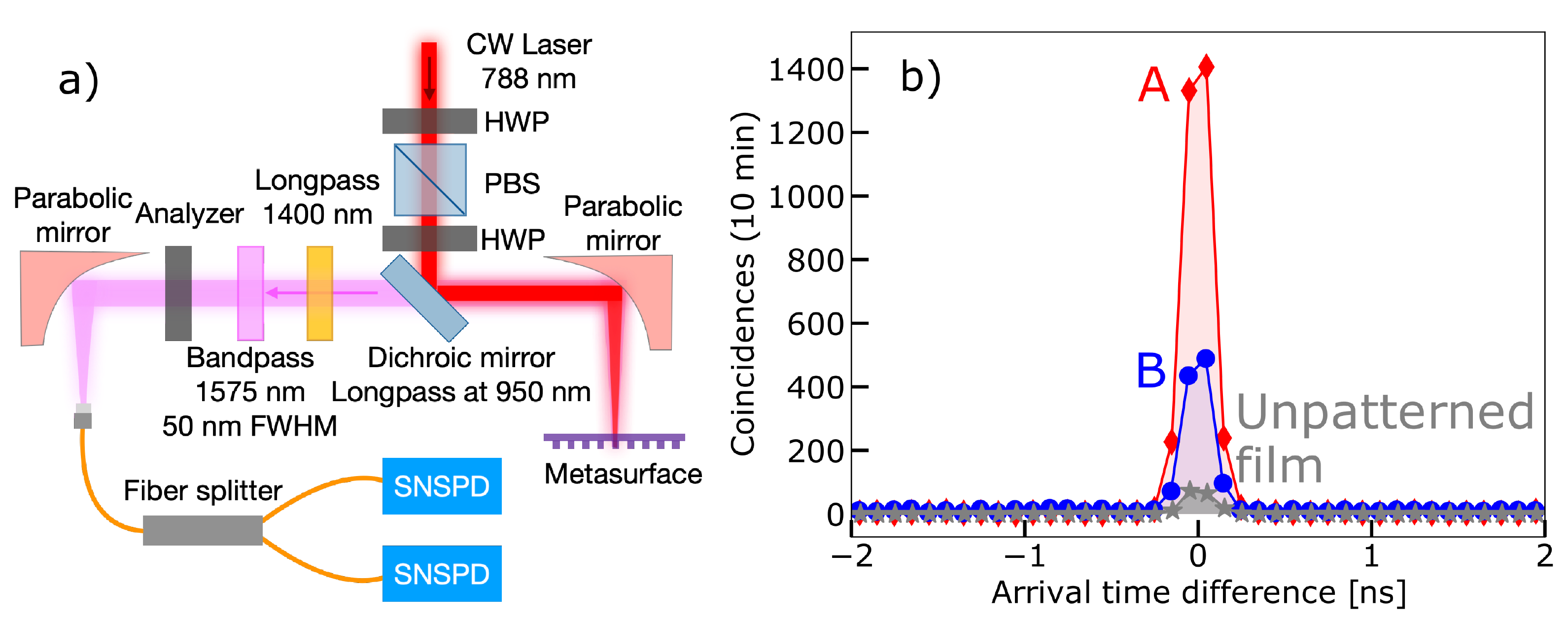}
  \caption{(a) Correlation experiment. A parabolic mirror focuses a cw pump into the QOM and collects backward-emitted SPDC. A dichroic mirror separates the SPDC radiation from the pump, and a 50\;nm FWHM bandpass filter centered at 1575\;nm transmits  nearly degenerate photon pairs. Another parabolic mirror feeds the SPDC into a Hanbury Brown -- Twiss  setup formed by a fiber splitter and two superconducting nanowire single-photon detectors (SNSPD). (b) Coincidence histograms of degenerate SPDC from QOMs A and B, shown by red diamonds and blue circles, respectively. The lines are guides to the eye. Gray stars show the coincidence histogram from an unpatterned LN film of the same thickness as the nanoresonators. 
  In all measurements the pump power is $ \sim 70\; \mathrm{mW}$ and the acquisition time $10$ min.}
  \label{fgr:histograms}
\end{figure}

We pumped the QOMs from the substrate side with a continuous-wave (cw) laser at powers of several tens of mW (Fig.~\ref{fgr:histograms}(a)). 
The pump laser was weakly focused using a parabolic mirror, resulting in a pump beam diameter of  $6 \; \mathrm{\mu m}$ on the QOM. The photon pairs generated in the backward direction were collected using the same parabolic mirror. The filtering system comprised a longpass filter for pump rejection, a bandpass filter of 50~nm full width at half maximum (FWHM) bandwidth, centered at 1575~nm, and a polarization analyzer selecting polarization along the LN optics axis. Finally, the generated photons were coupled into a fiber and registered by two single-photon detectors in a Hanbury Brown-Twiss (HBT) setup (see Fig.~\ref{fgr:histograms}(a) and Methods for more details). The collection numerical aperture (NA), determined by the fiber, was $0.14$. Single-photon detection events from the two detectors were analyzed using a time-to-digital converter. 

Figure~\ref{fgr:histograms}(b) shows coincidence histograms, i.e.\ the numbers of two-photon detection events versus the difference in the photon arrival times, measured for $\sim 70 \, \mathrm{mW} $ pump power over 10 min acquisition time, with the pump  polarized along the LN optic axis. 
%(x-coordinate). 
The red diamonds correspond to QOM A and the blue circles to QOM B. The peak in the middle indicates the simultaneous arrival of photons forming a pair. The coincidences-to-accidentals (peak-to-background) ratio (CAR) considerably exceeds $2$ in both cases, which clearly proves the generation of photon pairs in each QOM~\cite{Santiago2020}. The maximal obtained CAR is $361$. The rates of real coincidences, found from the total number of coincidences after subtraction of the accidental coincidences, are $5.4 \pm 0.1 \; \mathrm{Hz}$ and $1.8 \pm 0.1 \; \mathrm{Hz}$ for QOMs A and B, respectively.  The width of the coincidence histograms is given by the timing jitter of the detectors, which is about $180 \; \mathrm{ps}$.

We compared the photon-pair rates from the QOMs to that of an unpatterned LN film of the same thickness under the same experimental conditions (coincidence histogram shown by gray stars in Fig.~\ref{fgr:histograms}(b)). The peak values of the histograms measured in QOMs A, B are, respectively, $20$ and $7$ times higher than for the unpatterned LN. As shown below, we observe even a stronger enhancement of the pair generation rate by looking into a narrower spectral range. 	

\begin{figure}[ht!]
     \includegraphics[width=0.5\textwidth]{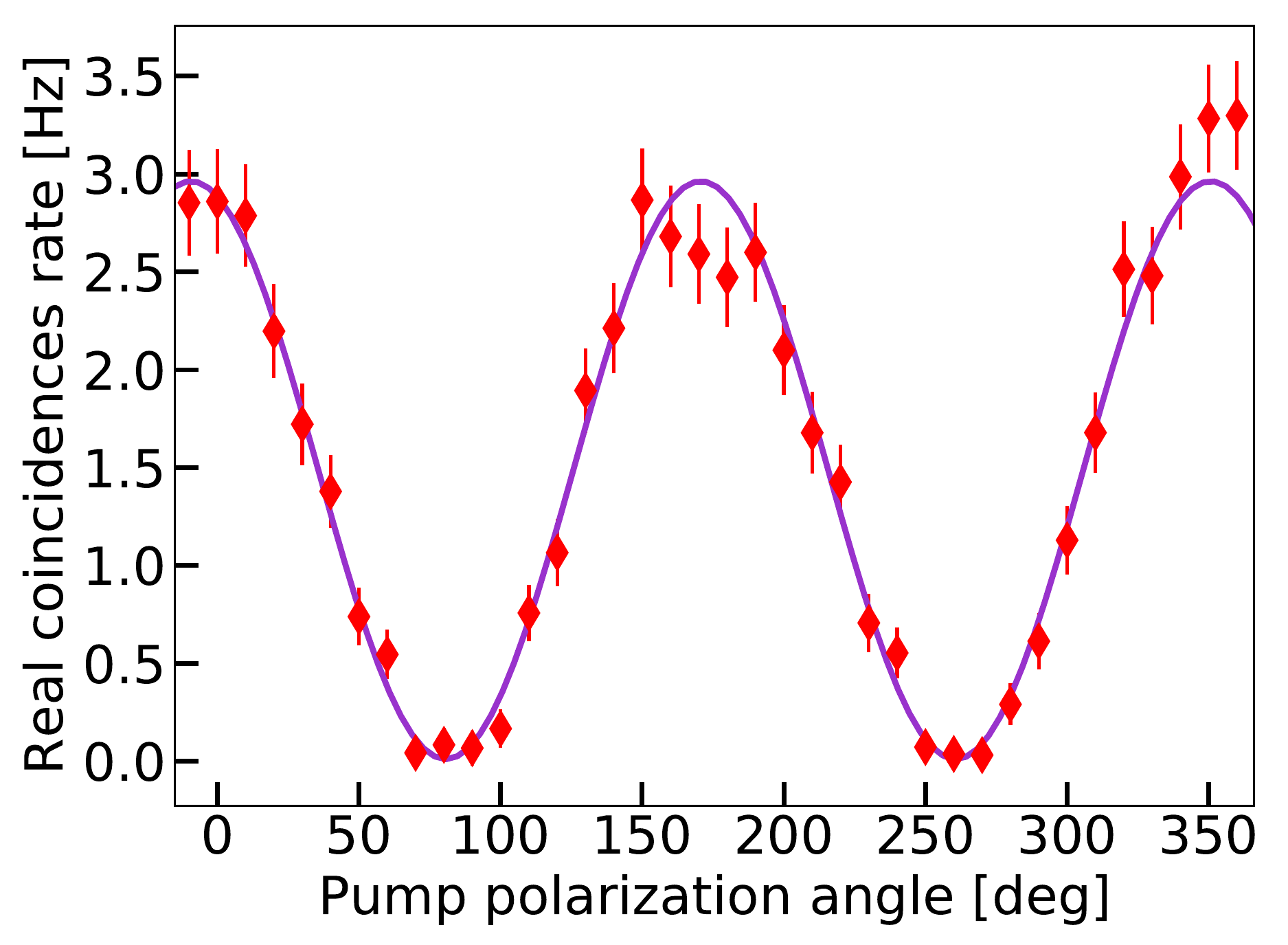}
  \caption{Real coincidence rate measured in  QOM A versus the pump polarisation angle with respect to the LN optic axis. The pump power was $\sim 50\; \mathrm{mW}$. The purple curve shows the theoretical cosine-squared dependence.}
  \label{fgr:polarization}
\end{figure}

To investigate the polarization dependence of SPDC, we have measured the coincidence rate while rotating the pump polarization angle and keeping the analyzer for signal and idler photons (Fig.~\ref{fgr:histograms}(a)) parallel to the LN optic axis. The result for QOM A, at the pump power $\sim 50 \; \mathrm{mW}$, is shown in Fig.~\ref{fgr:polarization}. The pair-generation rate scaled as $\cos^2 \theta$ (purple curve), $\theta$ being the pump polarization angle with respect to the LN optic axis. For the analyzer oriented orthogonally to the LN optic axis, no photon pairs could be registered regardless of the pump polarization. This behavior indicates that SPDC was indeed mediated by the $\chi^{(2)}_{zzz}$ tensor component.

Generating entangled photons with tailored spectral correlations is one of the major tasks in quantum optics. To demonstrate the influence of the resonance on the observed enhancement of SPDC and its spectral distribution, we measured the spectrum of the backward-emitted photon pairs via single-photon spectroscopy (SPS)~\cite{Valencia2002}. In SPS, the two-photon wavepacket is spread in time in a dispersive medium, in our case a 1\;km-long single-mode fiber (Corning SMF-28), and then coincidence events between two detectors are registered in a HBT setup (see Methods for more details). Due to the dispersion, the arrival time difference between signal and idler photons can be mapped to their spectral separation, and the spectrum can be retrieved from the coincidence histogram~\cite{Okoth2019,Santiago2020}. Note that the bandpass filter (see Fig.~\ref{fgr:histograms}(a)) was removed in these experiments . 

\begin{figure}[t!]
     \includegraphics[width=15.2cm]{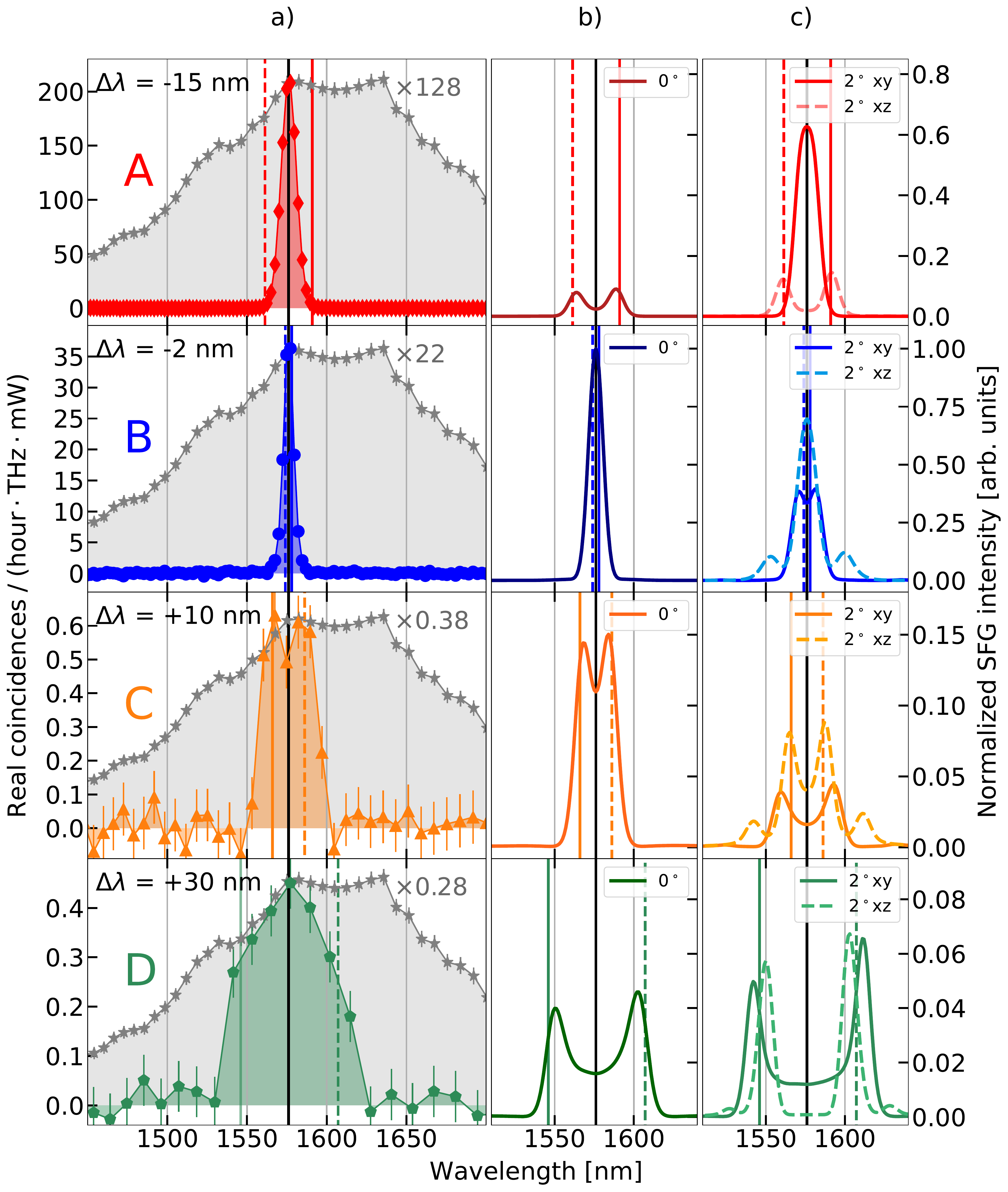}
  \caption{(a) Measured SPDC spectra from QOMs A (red), B (blue), C (orange) and D (green). Gray points show the SPDC spectrum from the unpatterned LN film. (b) The spectra obtained through the numerical simulation of SFG at normal incidence. (c) The SFG spectra calculated for the signal and idler incident at $\pm2^\circ$ to the normal direction in the $xy$ (solid lines) and $xz$ (dashed lines) planes.}  
  \label{fgr:SPDC_spectrum}
\end{figure}	  

The measured SPDC spectra for QOMs A, B, C, D are plotted in Fig.~\ref{fgr:SPDC_spectrum}(a). All spectra are symmetric around the wavelength of degenerate SPDC, $\lambda_{\text{deg}}=1576$~nm, which is because our experiment cannot distinguish between the signal and idler photons. The measured SPDC spectra are all well localized, in contrast to the spectrum of backward-emitted SPDC from the unpatterned LN film (gray points in all panels). Here, we observed a much larger bandwidth, limited only by the spectral sensitivity of our setup. The spectrum of the LN film is broad because the phase matching is relaxed~\cite{Santiago2020} and the vacuum field seeds SPDC uniformly over the spectrum~\cite{klyshkoNonlinearoptics}.

The width of the measured SPDC spectra strongly depends on the detuning $\Delta\lambda\equiv\lambda_{\text{deg}}-\lambda_{\text{ER}}$ between the degenerate wavelength $\lambda_{\text{deg}}$ and the wavelength $\lambda_{\text{ER}}$ of the electric resonance. The latter is marked in each panel by a solid vertical line. Furthermore, the conjugate wavelength of the electric resonance, $\lambda_{\text{conj}} = \left(\frac{2}{\lambda_{\text{deg}}}-\frac{1}{\lambda_{\text{ER}}}\right)^{-1}$, is marked with a dashed vertical line. This is the wavelength where the partner photon to the photon emitted at resonance is detected. We see that each measured spectrum is bounded by the electric resonance wavelength and its conjugate, whereas no photon pairs are observed at wavelengths corresponding to the magnetic resonance. This is due to the larger field enhancement of the electric resonances as well as the direction of their field, which is mainly along the LN optic axis and thus uses the $\chi^{(2)}_{zzz}$ tensor component~(Fig.~\ref{fgr:metasurface_design}(c)). The same effect was observed for classical frequency conversion in similar LN metasurfaces \cite{Fedotova2020}.

The most striking feature of the measured spectra is the giant enhancement of SPDC rate within a narrow resonance bandwidth. For QOM A, degenerate photon pairs are emitted at a rate 130 times higher than for the unstructured LN film. The moderate (20-times) advantage observed in Fig.~\ref{fgr:histograms}(b) resulted from averaging the spectrum over the 50-nm bandwidth of the filter, while the resonance is not broader than $10$ nm.

One might expect the enhancement to be stronger for QOMs with small detuning of the electric resonance from degeneracy
because efficient pair generation requires a high density of states at both signal and idler wavelengths~\cite{helt:2012:JOSAB}. While this is generally true, we see an additional tendency: the enhancement is stronger for metasurfaces with red-detuned resonances. For instance, although the resonance of QOM A is more detuned from degeneracy than the one of QOM B, it is more efficient, whereas QOM C is even less detuned, but very inefficient. 
This behavior can be explained as follows. According to our linear simulations and measurements (see the SI), the electric resonance gets blue-shifted under propagation at a nonzero angle to the metasurface normal. This is important for SPDC in ultrathin sources where photon pairs are emitted within a broad angle~\cite{Okoth2020}. By collecting SPDC into NA=0.14 we have the resonances effectively blue-shifted by approximately $10$ nm.

To confirm these qualitative considerations, we performed numerical simulations of sum-frequency generation (SFG) for signal and idler plane waves incident on the sample from the substrate side. According to the quantum-classical correspondence~\cite{helt:2015:OL,lenzini:2018:LSA,marino:2019:O}, the SFG efficiency is proportional to the rate of SPDC into the same modes. To obtain a full SFG
spectrum, this calculation was performed for signal and idler wavelengths satisfying energy conservation (see Methods for more details). In Fig.~\ref{fgr:SPDC_spectrum}(b) we plot the normalized intensity of SFG emitted in the backward direction (reversed geometry to the one shown in Fig.~\ref{fgr:metasurface_design}(a) for SPDC).

The simulations confirm the absence of the magnetic resonance contribution and the dependence of the spectral width on the resonance detuning.  
Moreover, additional simulations for SFG from signal and idler incident at moderate angles of $\pm2^\circ$ (Fig.~\ref{fgr:SPDC_spectrum}(c)) confirm the blue shift of the resonance. For QOM A, with detuning $\Delta\lambda = -15$~nm, the angular tilt leads to a narrower spectrum with increased count rate, as the resonance is shifting closer to degeneracy. This is in agreement with the experimental spectrum for QOM A. On the other hand, for QOM C, with $\Delta\lambda = +10$~nm, the angular tilt further increases the detuning from degeneracy, again in accordance with the measurement, where we observe a somewhat wider spectrum than expected from the normal-incidence resonance wavelengths. This blue shift also explains why QOM B with $\Delta\lambda=-2$~nm showed a lower count rate than QOM A with $\Delta\lambda=-15$~nm. Due to the collection of photons with tilted incidence and the resulting blue shift of the effective resonance wavelength, the latter gets closer to degeneracy for QOM A than for QOM B. For QOM D the simulations show good agreement of the spectrum bandwidth, but not the shape. We attribute this to the measurement instabilities: this is the QOM with the weakest SPDC signal, and the measurement was performed over 1 week when the sample might have been slightly displaced and ambient conditions might have changed.

To conclude, we observed photon pairs generated via SPDC in resonant metasurfaces. Importantly, our experiment was the first-time observation of {pronounced (CAR~$>2$) two-photon coincidences for SPDC in the reflection geometry. The photon-pair generation was strongly enhanced by the electric resonance: within its narrow bandwidth, enhancement by a factor of 130 was measured. Furthermore, the spectral width of the emitted photon pairs could be controlled through the detuning between the electric resonance and the SPDC degeneracy wavelength, although the generation efficiency decreases with increasing detuning.

The actual enhancement of SPDC in resonant QOMs is even higher if we take into account the pump diffraction. Since the QOM period is larger than the pump wavelength, diffraction into the first orders occurs, the zeroth order carrying  only a fraction of the incoming  pump power~\cite{Fedotova2020}.  
Photon pairs can be generated by each diffraction order of the pump; however, with our NA we only collect pairs corresponding to the zeroth order. We estimate that the measured photon-pair rate could be considerably increased by using optics with higher NA or modifying the QOM design.
Furthermore, the resonant modes of the QOM could be optimized to emit generated photon pairs exclusively in the forward or backward direction, thus enabling more efficient SPDC in either reflection or transmission geometries \cite{nikolaeva:2020:arxiv}. 

Our results are a first step towards the use of nonlinear metasurfaces as versatile sources of photon pairs. Apart from the spectral control that we demonstrated, QOMs will also enable far-reaching control of the spatial properties of SPDC, leading to unprecedented possibilities for the creation of complex two-photon quantum states.

\section{Methods}

\subsection{Correlation experiment in reflection}
    %We pumped the nonlinear metasurfaces with a cw pig-tailed diode laser delivering up to $\sim 70 \, \mathrm{mW} $ at 788\;nm.  
    The pump power was controlled using a half-wave plate (HWP) and a polarizing beam-splitter (PBS). We used another HWP to rotate the pump polarization. The pump beam with a diameter of $\sim 2.5\, \mathrm{mm}$ was focused into the metasurfaces using a $90 ^{\circ}$ off-axis gold-coated parabolic mirror with $15 \, \mathrm{mm}$  reflective focal length. The same parabolic mirror collected the backward-generated SPDC radiation. The parabolic mirror played an essential role as, being free from chromatic aberrations, it ensured the collection of photon pairs from the same point where the pump laser was focused. 

    A longpass dichroic mirror with cut-on wavelength at 950\;nm split the SPDC radiation and the portion of the pump laser reflected at the sample surface, with the SPDC being transmitted. Another longpass filter with cut-on wavelength at 1400\;nm filtered out the remaining pump and removed most part of the fluorescence. We detected degenerate photon pairs after a  50 nm FWHM bandpass filter centered at 1575 nm. A broadband linear film polarizer oriented parallel to the LN optic axis selected  photon pairs with $z$ polarization. The polarizer also served as an additional filter removing fluorescence. 

    A parabolic mirror identical to the one used for focusing and collection fed the radiation into the input facet of a 50:50 broadband single-mode fiber splitter (1550 nm $\pm$ 100 nm). Due to the identity of the two parabolic mirrors, the NA of the fiber ($0.14$) also determined the collection angle of SPDC. The two outputs of the fiber splitter were sent to infrared superconducting nanowire single-photon detectors (SNSPD). We registered the arrival time differences between the two detectors using a Swabian Instruments time-tagger (not shown).

\subsection{Single-photon spectroscopy}

    To measure the SPDC spectrum, we slightly modified  the setup described above. First, we inserted a 1\;km long single-mode fibre (Corning SMF-28) at the input facet of the fiber splitter so as to spread the biphoton wavepacket in time. Second, we removed the bandpass filter so that we could detect a larger bandwidth, from $1400$ nm to $1800$ nm. We overcame chromatic aberrations with the use of parabolic mirrors. Histograms of arrival times differences were recorded in a similar fashion as in the previous experiment.
    
   Because of the dispersion in the 1\;km long fiber, the arrival time differences were redistributed over a larger time interval. The delay $\tau$  between the signal and idler photons is given by 
   \begin{equation}
   \label{eqn:delay}
     \tau = L \cdot D(\lambda_{\text{deg}}) \cdot 2\pi c (\omega_{\text{s}}^{-1} - \omega_{\text{i}}^{-1}), 
   \end{equation}
   where  $L$  is the length of the fiber, $D(\lambda_{\text{deg}})$ is the dispersion parameter [in ps/(nm\(\cdot\)km)] of the fiber at the degenerate wavelength, and  the last factor at the right-hand side of Eq.~(\ref{eqn:delay}) is the wavelength separation between the signal and idler photons. Due to the energy conservation condition, Eq.~(\ref{eqn:delay}) imposes a one-to-one correspondence between $\tau$ and the wavelength of the signal photon. By inverting Eq.~(\ref{eqn:delay}) we map $\tau$, measured by the time tagger device, to the wavelength separation, from which the wavelength of the signal photon can be retrieved. In SMF-28 fiber, $D(\lambda_{\text{deg}} = 1576 \; \mathrm{nm}) \sim 18.8 \; \mathrm{ps/(nm \cdot  km)}$.~\cite{Zhu2015,Corning}  From Eq.~(\ref{eqn:delay}), the optical resolution of the system is 8.8~nm, as given by the timing jitter of the detectors and the length of the fiber.

\subsection{Simulations of sum-frequency generation}
    Simulations of sum-frequency generation were done in COMSOL Multiphysics using the undepleted pump approximation, and included three steps. The first two are linear simulations of electromagnetic field for a plane wave excitation at (1) the signal 
    frequency $\omega_{s}$ and (2) the corresponding idler frequency $\omega_{i}=\omega_{p}-\omega_{s}$, where $\omega_{p}$ is the frequency of our pump laser. Based on the electric fields from the first two steps, we calculated the nonlinear polarization (see Eq.~\ref{eqn:polarization}) inside a LN nanoresonator (pyramid and residual layer, see SI) which in turn served as a source for the final SFG simulation. This algorithm was repeated by varying the signal frequency 
    %from 1500~nm to 1650~nm
    and setting the idler frequency accordingly. %As the SFG wavelength of $788$~nm is smaller than the metasurfaces period the generated SFG has 
    For the simulations of SFG with oblique incidence of signal and idler the excitation at the first step (signal study) was tilted by $2^\circ$ and at the second step (idler study) by $-2^\circ$, both in $xy$ and $xz$ planes (see SI). The components of the nonlinear polarization had the form
    %\begin{equation}
    \begin{align}
    P_x^{NL}(\omega_p) &= 2\varepsilon_0 \left(d_{31}[ E_x(\omega_s) E_z(\omega_i) + E_z(\omega_s) E_x(\omega_i) ] - d_{22}[ E_x(\omega_s) E_y(\omega_i)+E_y(\omega_s) E_x(\omega_i) ]\right),\nonumber\\
    P_y^{NL}(\omega_p) &= 2\varepsilon_0 \left(d_{22}[E_y(\omega_s) E_y(\omega_i) -E_x(\omega_s) E_x(\omega_i)] + d_{31}[E_y(\omega_s) E_z(\omega_i)+E_z(\omega_s) E_y(\omega_i)]\right),\nonumber\\
    P_z^{NL}(\omega_p) &= 2\varepsilon_0 \left(d_{31}[ E_x(\omega_s) E_x(\omega_i)+E_y(\omega_s) E_y(\omega_i) ] + d_{33}E_z(\omega_s) E_z(\omega_i)\right),%\nonumber
    \label{eqn:polarization}
    \end{align}
    %\end{equation}
    where %$P_i^{NL}(\omega_p)$ was the $i$-component of nonlinear polarization, 
    $E_{x,y,z}(\omega)$ were the components of the signal ($\omega_s$) or idler ($\omega_i$) electric field, and $d_{22}=1.9\:\text{pm/V},d_{31}=-3.2\:\text{pm/V},d_{33}=-19.5\:\text{pm/V}$ at 1313~nm.\cite{shoji1997}

    The resulting SFG spectra were multiplied by the SPDC spectrum from the wafer (gray stars in Fig.~\ref{fgr:SPDC_spectrum}(a)), to take into account the detectors sensitivity, and convolved with a Gaussian with FWHM 8.8~nm, to take into account the detectors timing jitter. 

\begin{acknowledgement}

T. S. C. and M. A. W. are part of the Max Planck School of Photonics supported by BMBF, Max Planck Society, and Fraunhofer Society.
This work was supported by DAAD GSSP-2017, the German Research Foundation (STA 1426/2-1, PE 1524/13-1, IRTG 2101, SE 2749/1-1, 398816777-SFB1375), the Federal Ministry of Education and Research (13N14147, 03ZZ0434, 13N14877), the Fraunhofer Society within the graduate school of the Leistungszentrum Photonik, and the Thuringian State Government within its ProExcellence initiative (ACP2020).

\end{acknowledgement}

\begin{suppinfo}
The Supporting Information is available free of charge via the internet at http://pubs.acs.org.
It contains the linear spectra of selected metasurfaces in different geometries (reflection and transmission) for normal and oblique incidence, the multipole decompositions, and the details on the field enhancement.

\end{suppinfo}

\bibliography{references}

\end{document}